\documentclass[tighten,twocolumn]{aastex63}

\usepackage{graphicx}
\usepackage{ulem}
\usepackage{amsmath}
\usepackage{wrapfig}
\usepackage[thinlines]{easytable}
\usepackage{tikz}
\usepackage{hyperref}

\graphicspath{{./}{figures/}}

\begin{document}
\shorttitle{}
\shortauthors{Kremer et al.}

\title{Can slow pulsars in Milky Way globular clusters form via partial recycling?}

\correspondingauthor{Kyle Kremer}
\email{kykremer@ucsd.edu}

\author[0000-0002-4086-3180]{Kyle Kremer}
\affiliation{Department of Astronomy \& Astrophysics, University of California, San Diego; La Jolla, CA 92093, USA}

\author[0000-0001-9582-881X]{Claire S.\ Ye}
\affiliation{Canadian Institute for Theoretical Astrophysics, University of Toronto, 60 St. George Street, Toronto, Ontario M5S 3H8, Canada}

\author[0000-0003-3944-6109]{Craig O.\ Heinke}
\affiliation{Department of Physics, University of Alberta, CCIS 4-183, Edmonton, AB, T6G 2E1, Canada}

\author[0000-0001-6806-0673]{Anthony L. Piro}
\affiliation{The Observatories of the Carnegie Institution for Science, Pasadena, CA 91101, USA}

\author[0000-0001-5799-9714]{Scott M.\ Ransom}
\affiliation{NRAO, 520 Edgemont Road, Charlottesville, VA 22903, USA}

\author[0000-0002-7132-418X]{Frederic A.\ Rasio}
\affiliation{Center for Interdisciplinary Exploration \& Research in Astrophysics (CIERA) and Department of Physics \& Astronomy \\ Northwestern University, Evanston, IL 60208, USA}

\begin{abstract}
Alongside the population of several hundred radio millisecond pulsars currently known in Milky Way globular clusters, a subset of six slowly spinning pulsars (spin periods $0.3-4$\,s) are also observed. With inferred magnetic fields $\gtrsim 10^{11}\,$G and characteristic ages $\lesssim10^8\,$yr, explaining the formation of these apparently young pulsars in old stellar populations poses a major challenge. One popular explanation is that these objects are not actually young but instead have been partially spun up via accretion from a binary companion. In this scenario, accretion in a typical low-mass X-ray binary is interrupted by a dynamical encounter with a neighboring object in the cluster. Instead of complete spin up to millisecond spin periods, the accretion is halted prematurely, leaving behind a ``partially recycled'' neutron star. In this Letter, we use a combination of analytic arguments motivated by low-mass X-ray binary evolution and $N$-body simulations to show that this partial-recycling mechanism is not viable. Realistic globular clusters are not sufficiently dense to interrupt mass transfer on the short timescales required to achieve such slow spin periods. We argue that collapse of massive white dwarfs and/or neutron star collisions are more promising ways to form slow pulsars in old globular clusters.
\vspace{1cm}
\end{abstract}

\section{Introduction}
\label{sec:intro}

Since the initial discoveries in the 1980s \citep{Lyne1987}, a population of over 300 radio millisecond pulsars (MSPs) has now been uncovered in Milky Way globular clusters.\footnote{For an up-to-date list, see Paulo Freire's \hyperlink{https://www3.mpifr-bonn.mpg.de/staff/pfreire/GCpsr.html}{``Pulsars in Globular Clusters''} catalog.}
The canonical method for forming a MSP in a cluster is via mass transfer in a low-mass X-ray binary \citep[LMXB;][]{Alpar1982}. Here, angular momentum is deposited onto the neutron star via an accretion disk, spinning up the neutron star to spin periods as short as a few milliseconds. This formation mechanism is corroborated by the detection of large numbers of LMXBs in globular clusters, both in luminous active states with luminosities in excess of $10^{36}\,$erg/s \citep{Clark1975} and in low-luminosity ($\lesssim10^{34}\,$erg/s) quiescent states \citep{Heinke2003}. The specific abundances (number per unit mass) of both MSPs and LMXBs in globular clusters are more than $100$ times higher than in the Galactic field \citep{Pooley2003,Bahramian2013}. This overabundance is understood to be a result of the high stellar densities of globular clusters, which enable dynamical formation pathways for these systems not accessible for isolated binaries \citep[e.g.,][]{Fabian1975,Heggie1975,RasioShapiro1991,Davies1992,Ivanonva2008,Ye2019}.

The full population of radio pulsars in globular clusters is diverse, including both single MSPs and binary systems of several varieties: standard low-mass MSP binaries with white dwarf companions of mass $0.1-0.2\,M_{\odot}$ and orbital periods of days or more \citep[formed via the classic LMXB scenario involving mass transfer from a (sub)giant donor;][]{Tauris1999}; ``black widow'' pulsars with very low mass companions ($M_c \lesssim 0.04\,M_{\odot}$) and orbital periods of hours \citep{Fruchter1988}, and ``redback'' pulsars with $\gtrsim 0.1\,M_{\odot}$ main sequence-like companions that exhibit irregular eclipses and erratic timing \citep{Strader2019}. The MSP binary fraction varies significantly across different cluster types. In low-concentration globular clusters with $r_c/r_h > 0.2$ \citep[$r_c$ and $r_h$ are the core and half-light radius, respectively;][]{Harris1996}, roughly 33\% of MSPs are single. This is comparable to the fraction of isolated MSPs in the Galactic field -- 36\% for pulsars with spin periods less than $30\,$ms \citep{Manchester2005} -- hinting that the majority of single MSPs in low-density GCs became single via evolutionary mechanisms similar to those operating for isolated binaries \citep[e.g.,][]{Bildsten2002}. Meanwhile, the singles fraction increases to 57\% in denser clusters with $r_c/r_h < 0.2$ and up to 83\% in the ultra-dense core-collapsed clusters \citep{Ye2024}. This suggests that as cluster density increases, additional dynamically-induced mechanisms for forming single MSPs become important. This may include disruption of MSP binaries via dynamical exchange encounters \citep{SigurdssonPhinney1995} or possibly more exotic mechanisms including accretion following a stellar collision \citep{Davies1992,Kremer2022,Ye2024}.

The vast majority of cluster pulsars have spin periods less than $30\,$ms and low magnetic fields consistent with expectations for the standard recycling scenario \citep[e.g.,][]{PhinneyKulkarni1994}. However, a subpopulation of \textit{slowly spinning} pulsars with spin periods of $0.3-1\,$s and inferred magnetic fields of roughly $10^{11}-10^{12}\,$G has also been found
\citep{Lyne1996,Boyles2011}. In the past year, two additional slow pulsars with spin periods of roughly $2$ and $4\,$s have been identified in M15 by the Five-hundred-meter Aperture Spherical Telescope \citep[FAST;][]{Zhou2024,Wu2024}, bringing the current total up to six. With characteristic ages of $10^8\,$yr or less, these apparently young pulsars are impossible to explain via standard massive-star evolution, as the $\lesssim 50\,$Myr lifetime of any plausible massive-star progenitor is significantly less than the typical $10\,$Gyr age of their globular cluster hosts. Alternative formation channels have been proposed, in particular involving collapse of massive white dwarfs triggered by accretion from a binary companion \citep{Tauris2013} or formed through a binary merger \citep{Kremer2023}.

Another possibility is that these six slowly spinning pulsars are not actually young but instead have been partially recycled \citep{VerbuntFreire2014}. Binaries in globular clusters experience frequent dynamical encounters with other stars and binaries \citep{HeggieHut2003}. If a neutron star LMXB is interrupted mid-accretion by an interloping object, the spin-up process may be halted before millisecond spin periods can be achieved, leaving behind a partially recycled pulsar with a relatively large spin period and magnetic field. This process would be most common in the densest globular clusters, where dynamical encounters capable of interrupting mass transfer are most frequent. Notably, \textit{all six} of these slow pulsars are found in exceptionally dense clusters that have (or have nearly) undergone cluster core collapse \citep{Kremer2023}, providing tantalizing evidence for the partial recycling process \citep{VerbuntFreire2014}.

Here we examine in further detail whether the partial recycling scenario is indeed a viable mechanism for explaining these six slow and apparently young pulsars. In Sections~\ref{sec:LMXB} and \ref{sec:lower_limit} we present an order-of-magnitude style estimate based on LMXB evolution and observations and in Section~\ref{sec:CMC} we explore this scenario using $N$-body models representative of the core-collapsed globular clusters that host these objects. In both cases, we conclude that the partially recycling scenario is not viable for typical LMXBs. In Section~\ref{sec:alternatives}, we discuss two alternative formation scenarios. We summarize our results in Section~\ref{sec:conclusion}.

\section{Comparing timescales for spin-up and dynamical encounters}
\label{sec:LMXB}

\begin{figure*}
    \centering
    \includegraphics[width=0.85\textwidth]{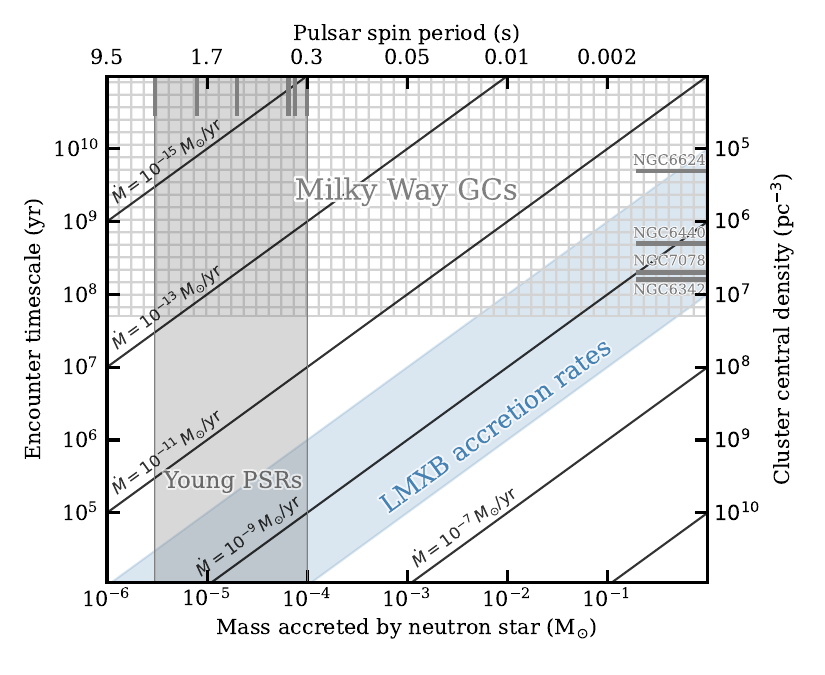}
    \caption{Schematic illustration of the spin-up timescale for a neutron star accreting in an LMXB (solid black lines show various accretion rates) compared to the dynamical encounter timescales in typical globular clusters. The vertical gray band indicates the range of spin periods for the six slowly spinning radio pulsars known in Milky Way globular clusters \citep{Boyles2011, Zhou2024}, with the gray ticks showing specific spin period values. The diagonal blue band shows accretion rates expected at onset of mass transfer for typical LMXBs. The intersection of the gray and blue bands indicates where ``partial recycling'' may produce slow pulsars via dynamical interruption of the spin-up process \citep{VerbuntFreire2014}. The hatched gray region displays central densities (and associated encounter timescales from Equation~\ref{eq:t_enc}) for Milky Way globular clusters. As shown by the lack of overlap between the blue, gray, and hatched regions, realistic globular clusters do not have sufficiently high central densities for the partial recycling scenario to successfully operate. We propose alternative scenarios in Section~\ref{sec:alternatives}.}
    \label{fig:partial_recycling}
\end{figure*}

In an LMXB, mass transferred from the donor deposits angular momentum which spins up the neutron star. For details of the accretion process, see \citet{Pringle1972,GhoshLamb1979,Frank2002,Rappaport2004}. The mass that must be accreted in order to spin up a neutron star to (equilibrium) spin period $P$ can be estimated as

\begin{equation}
    \label{eq:spin_up}
    \Delta M \approx 0.05\, f \Bigg( \frac{P}{3\,\rm{ms}} \Bigg)^{-4/3} M_{\odot}
\end{equation}
\citep[e.g.,][]{PhinneyKulkarni1994}, where $f \gtrsim 1$ is a factor that depends on the accretion details. 
We adopt the widely-accepted idea that the magnetic field decays rapidly in the early phases of the accretion process \citep{Taam1986,Shibazaki1989,vandenHeuvel1995,Tauris2012}, so that the Alfv\'{e}n radius is comparable to the neutron star radius throughout most of the spin-up process. Assuming a constant mass accretion rate $\dot{M}$ (already a conservative assumption; see Section~\ref{sec:lower_limit}) and $f\approx 1$, the time to spin up a neutron star to spin period $P$ is roughly

\begin{equation}
    \label{eq:t_spinup}
    \tau_P \approx \Delta M/\dot{M} \approx 2\times 10^7 \Bigg( \frac{P}{3\,\rm{ms}}\Bigg)^{-4/3} \Bigg( \frac{\dot{M}}{0.1\dot{M}_{\rm Edd}}\Bigg)^{-1} \,\rm{yr},
\end{equation}
where we have scaled to the Eddington accretion rate $\dot{M}_{\rm Edd} \approx 2\times10^{-8}\,M_{\odot}/$yr for a neutron star.

LMXBs with mass transfer driven by nuclear evolution of a (sub)giant companion pass through an initial phase of high $\dot{M}$ up to $10^{-9}-10^{-8}\,M_{\odot}/$yr \citep[e.g.,][]{Webbink1983,Rappaport1995,Tauris1999,Tauris2012}. Donors on the main sequence (masses between $0.2-1\,M_{\odot}$) with mass transfer driven by gravitational radiation and/or magnetic braking may lead to slightly smaller accretion rates roughly $10^{-10}\,M_{\odot}/$yr \citep[e.g.,][]{Verbunt1993}. For $\dot{M} \gtrsim 10^{-10}\,M_{\odot}/$yr, the accretion time to reach $P\approx 1\,$s is less than roughly $2\times10^5\,$yr (Equation~\ref{eq:t_spinup}). Thus, in order for the partial recycling scenario to provide a viable explanation for our six slowly spinning pulsars, dynamical encounter timescales of order $10^5\,$yr after the onset of mass transfer are required. In the case of such an encounter, the mass transfer could in principle be halted before millisecond spin periods are attained.

The orbital separation for a $1.2\,M_{\odot}$ neutron star with a $0.9\,M_{\odot}$ subgiant companion at the onset of Roche lobe overflow is $\lesssim 10R_{\odot}$. The typical encounter timescale for such a binary in the center of a core-collapsed globular cluster like M15 is:

\begin{multline}
    \label{eq:t_enc}
    \it{t}_{\rm enc} = \Bigg[n \pi a^2 \Bigg(1+\frac{2GM_{\rm tot}}{a \sigma_v^2} \Bigg)\sigma_v \Bigg]^{-1}  \\ \approx 10^{8} \Big( \frac{n}{10^7\,\rm{pc}^{-3}} \Big)^{-1} \Big( \frac{a}{10\,R_{\odot}} \Big)^{-1} \Big( \frac{M_{\rm tot}}{2\,M_{\odot}} \Big)^{-1} \Big( \frac{\sigma_v}{10\,\rm{km/s}} \Big)\,\rm{yr},
\end{multline}
where the simple scaling in the second line arises because gravitational focusing is dominant for these very hard binaries \citep[e.g.,][]{BinneyTremaine2008}. For $P\approx 1\,$s and $\dot{M}\gtrsim10^{-10}\,M_{\odot}/$yr, this encounter timescale exceeds $\tau_P$ by  factor of nearly $10^3$. This suggests that interruption of the spin-up process via dynamical encounters is a major challenge in typical globular clusters.

We demonstrate this point schematically in Figure~\ref{fig:partial_recycling}. The horizontal axes show accreted mass (bottom) and corresponding spin period via accretion spin up (top) from Equation~\ref{eq:spin_up}. The vertical gray band shows the range of spin periods for the six slow pulsars in Milky Way globular clusters, with the specific spin values marked as gray ticks on the top axis. The vertical axes show cluster central density (right) and corresponding encounter timescale (left) for an LMXB at Roche lobe overflow (Equation~\ref{eq:t_enc}). The hatched horizontal region shows the range of central densities (including both luminous and nonluminous stars) of Milky Way globular clusters from \citet{BaumgardtHilker2018}. We also show as gray ticks the central densities for the four specific globular clusters that host slow pulsars: NGC~6624 (two sources), NGC~6440, NGC~7078 (two sources), and NGC~6342. The diagonal black lines show several time-averaged accretion rate values, assuming that $M_{\rm acc}$ is accreted at a roughly constant rate over a time $t_{\rm enc}$. The blue band indicates ``allowed'' values of $\dot{M} \in [10^{-10},10^{-8}]M_{\odot}/\rm{yr}$, as inferred from LMXB evolutionary arguments discussed above (we discuss the lower limit on $\dot{M}$ in further detail in Section~\ref{sec:lower_limit}).

The intersection of the blue and gray bands marks the region of parameter space where a partial recycling scenario is viable. Here, the dynamical encounter timescale is sufficiently short to potentially interrupt accretion before significant spin up can occur. However, as shown, for central densities similar to Milky Way globular clusters, such short timescales are not expected. For $\dot{M} \gtrsim 10^{-10}M_{\odot}/\rm{yr}$, the spin up time to $300\,$ms (the fastest spinning of our six slow pulsars) is $\lesssim 10^6\,$yr, nearly $10^2$ times smaller than the typical encounter timescale in even the densest Galactic clusters and nearly $10^4$ times smaller than the encounter timescale expected in NGC~6624, which contains two slow pulsars. In short, Milky Way globular clusters are not dense enough to enable partial recycling of LMXBs. Any LMXB formed in a typical globular cluster will fully recycle its neutron star to millisecond spin periods before encountering another object.

\section{Further constraints on the minimum binary mass transfer rate}
\label{sec:lower_limit}

As shown in Figure~\ref{fig:partial_recycling}, the partial recycling scenario could plausibly operate in Galactic globular clusters if accretion rates in excess of $10^{-12}\,M_{\odot}/$yr can be avoided entirely. For (sub)giant or main-sequence donors, this is not possible. However, this could be possible for very low-mass ($M\lesssim 0.01\,M_{\odot}$) brown dwarf or white dwarf donors \citep[e.g.,][]{Howell2001,DeloyeBildsten2003,KingWijnands2006}. For isolated binaries, forming a neutron star binary with such low-mass stellar companions is a major evolutionary challenge \citep[e.g.,][]{Pfahl2003}. This challenge is likely exacerbated in globular clusters, as binary exchanges involving neutron stars tend to pair up more massive companions and swap out lower-mass objects \citep{SigurdssonPhinney1995}. Furthermore, even if such a binary could form in a cluster, the gravitational wave inspiral to reach contact \citep[$\gtrsim 10^8\,$yr for a $1.2M_{\odot}+0.01M_{\odot}$ binary with $a \gtrsim 0.2R_{\odot}$;][]{Peters1964} is longer than the spin up time once accretion begins (roughly $10^7\,$yr for $P\approx 1\,$s and $\dot{M} \approx 10^{-12}M_{\odot}/$yr; Equation~\ref{eq:t_spinup}), suggesting such a system is more likely to be disrupted \textit{prior} to onset of mass transfer than after.

Observations of accreting X-ray pulsars provide further insight into the relevant minimum $\dot{M}$. The only known LMXBs \citep[2A\,1822-371, 4U\,1626-67, GRO\,1744-28, Her\,X-1, and GX\,1+4;][]{PatrunoWatts2021} with measured spin periods above roughly $300\,$ms all have mass transfer rates of $10^{-10}-10^{-8}\,M_{\odot}$/yr \citep{ChakrabartyRoche1997,Chakrabarty1997,RappaportJoss1997,Coriat2012,GonzalezGalan2012,Heinke2013,BakNielsen2017}. The combination of high $\dot{M}$ and slow spins indicate these LMXBs have likely just started mass transfer. The absence of observed systems with both slow spins \textit{and} low accretion rates implies systems that avoid $\dot{M} \gtrsim 10^{-10}\,M_{\odot}$/yr entirely likely do not form, consistent with expectations from evolutionary calculations discussed in the previous section. 

Another interesting case is IGR J17480-2466 \citep{Bordas2010,Markwadt2010} which is found in Terzan 5, one of the densest Galactic globular clusters. This source has a slightly faster spin period of $90\,$ms, and inferred mass transfer rate of roughly $10^{-11}\,M_{\odot}$/yr \citep{Degenaar2011,Potekhin2019}. \citet{Patruno2012} demonstrates this binary likely entered its accretion phase within the past $10^5/\Delta\,$yr (where $\Delta$ is the unknown accretion duty cycle), also consistent with the picture that the initial spin-up phase at the onset of mass transfer occurs promptly, not leaving sufficient time for dynamical encounters that may interrupt the spin-up process. 

Very low binary mass transfer rates could potentially be produced via wind accretion. For main sequence donors, the predicted wind-loss rates \citep[roughly $10^{-16}-10^{-14}\,M_{\odot}/$yr;][]{MaccaronePatruno2013} would require timescales in excess of several Gyr to partially recycle a neutron star, even before accounting for a wind accretion efficiency likely well below the Bondi-Hoyle-Littleton rate \citep{Bleach2002}. Wind accretion from giants may be sufficient, however symbiotic stars are quite rare in globular clusters. Also these are expected to be most concentrated in the \textit{least} dense globular clusters \citep{Belloni2020} and thus would not replicate the clear observational preference for slow pulsars to be found in the densest clusters. In this case, wind accretion is likely not a viable mechanism.

As a final comment, we note that Equation~(\ref{eq:t_spinup}) is derived assuming a steady, continuous accretion rate. However, accreting neutron stars observed with mean mass transfer rates $\lesssim 10^{-10}\,M_{\odot}/$yr tend to accrete via episodic transient accretion (with much higher instantaneous accretion rates) owing to accretion disk instabilities. This can significantly reduce both $\Delta M$ and the timescale required to recycle neutron star spin \citep{Bhattacharyya2017,D'Angelo2017,Kar2024}, and thus $\tau_P$ in Equation~(\ref{eq:t_spinup}) can be viewed as an upper limit on the true accretion timescale. More precise treatment of this episodic accretion does not change our basic conclusions -- in fact, it makes the mismatch with the dynamical encounter timescale of Equation~(\ref{eq:t_enc}) even worse.

\section{Confirmation from $N$-body models}
\label{sec:CMC}

\begin{figure}
    \centering
    \includegraphics[width=\linewidth]{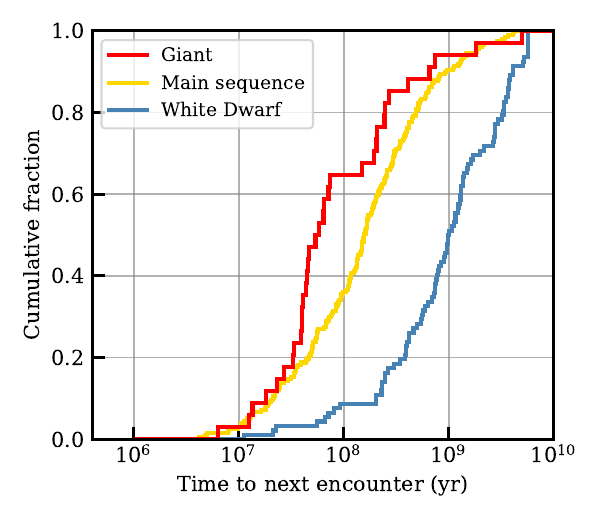}
    \caption{Distribution of time from the onset of mass transfer to the time of next dynamical encounter for all neutron star LMXBs formed in our \texttt{CMC} models. Different colors denote different donor types. As expected from Fig.~\ref{fig:partial_recycling}, we find that encounter timescales $<10^5\,$yr (required to partially recycle a neutron star) never occur.}
    \label{fig:CMC}
\end{figure}

As an additional test, we explore the evolution of LMXBs and neutron star spin up in a set of $N$-body models. We use the Monte Carlo code \texttt{CMC} \citep{Rodriguez2022}. For an up-to-date discussion of the treatment of formation and evolution of pulsars in \texttt{CMC}, see \citet{Ye2019,Ye2024}. Here we use a suite of ten \texttt{CMC} models: five models from \citet{Ye2024_massgap} and \citet{Ye2024} that closely match the core-collapsed cluster NGC~6752 and five new models that closely match the core-collapsed cluster NGC~6624 \citep{Rui2021}, host to two slow pulsars with spin periods of roughly $400\,$ms \citep{Boyles2011}.

Once an LMXB forms in \texttt{CMC} \citep[via dynamical exchange or binary evolution;][]{Ye2019}, the subsequent mass transfer is modeled with the binary evolution code \texttt{COSMIC} \citep{Breivik2020}, which allows the neutron star to be gradually spun up based upon the accretion rate from its companion \citep[in turn following the prescriptions of][]{Kiel2008}. As the LMXB evolves, \texttt{CMC} also computes the probability for the LMXB to undergo strong dynamical encounters with other nearby single stars and/or binaries
in the cluster \citep{Fregeau2007}. As such, \texttt{CMC} self-consistently computes the types of dynamical encounters that would potentially disrupt an accreting LMXB and produce a partially-recycled neutron star.

In Figure~\ref{fig:CMC}, we show the distribution of timescales between the onset of mass transfer for all neutron star LMXBs in our \texttt{CMC} models and the time of the next encounter. We separate encounters based on the stellar type of the donor star: white dwarf donor (blue), giant donors (red), and main-sequence donors (yellow). Since giant donors have the largest stellar radii, they fill their Roche lobes at the widest orbital separations. As a result, systems with giant donors have the largest cross section for encounter (see Equation~\ref{eq:t_enc}) and thus feature relatively short encounter timescales. Note that the median encounter timescales found in our \texttt{CMC} models are roughly comparable to the simple `$n\sigma v$' estimate of Equation~\ref{eq:t_enc}.

As discussed in Section~\ref{sec:LMXB}, the range of encounter timescales identified in our \texttt{CMC} models are much longer than the typical accretion times required to spin up a neutron star to $1\,$s spin periods. Not surprisingly, we identify \underline{\textit{zero}} cases in our \texttt{CMC} models of formation of a partially-recycled pulsar. Once binary mass transfer begins, all neutron stars in our models are spun up to millisecond spin periods before their next encounter. We conclude again that partial recycling of LMXBs does not offer a viable formation pathway for the six slowly spinning pulsars observed.

\section{Alternative explanations}
\label{sec:alternatives}

In lieu of the LMXB partial recycling scenario, we propose two alternative formation channels that both are strongly enhanced in core-collapsed clusters. 

\subsection{Massive white dwarf mergers}
\label{sec:wd}

Our favored model for forming young pulsars is collapse following the merger of a super-Chandrasekhar white dwarf binary, described in detail in \citet{Kremer2023}. The centers of old core-collapsed globular clusters are expected to harbor large populations of massive white dwarfs, as suggested by both simulations \citep[e.g.,][]{Kremer2021} and observations \citep[e.g.,][]{Vitral2022}. Within these clusters, dynamical interactions form massive white dwarf binaries, and ultimately mergers, at a rate of roughly $10^{-7}$ mergers per year in the full Milky Way cluster population \citep{Kremer2023}. More than $90\%$ of these white dwarf mergers have total mass in excess of the Chandrasekhar limit, which, along with their chemical compositions and mass ratios, make them strong candidates for collapse into neutron stars \citep{King2001,Dessart2006,Schwab2021}.

In \citet{Kremer2023} we argued, based in part upon spins and magnetic fields of observed white dwarf merger products in the Galactic field \citep{Ferrario2015,Caiazzo2021}, that these massive white dwarf mergers would collapse to pulsars with spin period $10-100\,$ms and magnetic fields $10^{11}-10^{13}\,$G. On timescales of roughly $10^8\,$yr (depending on the magnetic field strength at formation), these young pulsars will spin down via magnetic dipole radiation to $0.1-1\,$s spin periods before ultimately becoming undetectable as radio sources \citep{Ruderman1975}. The $10^8\,$yr lifetimes and white dwarf merger rate imply a population of roughly 10 young pulsars in the Milky Way globular clusters today, comparable to the six currently known. Furthermore, \citet{Kremer2023_m87} showed that the white dwarf merger rate is enhanced in core-collapsed clusters by at least a factor of $100$ relative to non-core-collapsed clusters, naturally accounting for the clear preference for finding young pulsars in only the densest systems. 

Accretion induced collapse of massive white dwarfs via stable mass transfer from binary companions may provide a similar pathway for formation of young neutron stars in globular clusters \citep{Tauris2013}, and likely operates at similar rate to mergers  \citep{Kremer2021_frb}. Although not in a globular cluster, the aforementioned 4U\,1626-67 source \citep{Giacconi1972} may exemplify this scenario. This system contains an apparently young ($P=7.66\,$s) accreting pulsar with a $10^{12}\,$G field in an ultracompact 42-minute binary with a $\lesssim 0.026\,M_{\odot}$ white dwarf donor \citep{Verbunt1990,Hemphill2021}. However, given that all but one of the six slow pulsars known in globular clusters are single, the original binary companion would need to be disrupted by a subsequent dynamical encounter \citep{Kremer2023}.

\subsection{Neutron star collisions}
\label{sec:NScollisions}

Stellar collisions are a common occurrence throughout the lifetimes of globular clusters \citep{HeggieHut2003}. If a previously inactive neutron star collides with a main-sequence or giant star, accretion onto the neutron star may spin up the neutron star and reactivate it as a radio source \citep{Davies1992,SigurdssonPhinney1995,HansenMurali1998,CamiloRasio2005}. This collision scenario was also pointed out by \citet{VerbuntFreire2014} as an additional partial recycling mechanism alongside interruption of an accreting LMXB.

The details of the accretion flow following the disruption of a star are highly uncertain and depend in part on the role of accretion feedback in unbinding material initially bound to the neutron star \citep[e.g.,][]{CamiloRasio2005} and on whether hypercritical accretion in excess of the Eddington limit can be achieved \citep[e.g.,][]{Chevalier1993,Fryer1996,BetheBrown1998,MacLeod2015}. In the specific cases where roughly $10^{-5}-10^{-4}M_{\odot}$ is ultimately accreted, spin periods of $0.1-1\,$s could be achieved via partial recycling. Using \texttt{CMC} cluster simulations, \citet{Kremer2022} estimated a neutron star collision rate of roughly $10^{-7}\,\rm{yr}^{-1}$ in the Milky Way clusters, comparable to the white dwarf merger rate of Section~\ref{sec:wd}. Again assuming the $P\approx 1\,$s pulsars formed via these collisions are observable for $10^8\,$yr as radio sources, the detection of roughly six sources at present in the Milky Way is plausible.

However this requires fine tuning. If less than $10^{-5}\,M_{\odot}$ is accreted, the recycling is negligible and if more is accreted, millisecond spin periods can be achieved. For example, \citet{Kremer2022} and \citet{Ye2024} argued these collisions may provide the most viable mechanism for the apparent overabundance of single MSPs in core-collapsed clusters discussed in Section~\ref{sec:intro}. Thus, although partial recycling via stellar collisions cannot be definitively ruled out as a pathway for forming slow pulsars, the white dwarf merger scenario may offer a simpler explanation.

\section{Conclusions and Discussion}
\label{sec:conclusion}

In addition to the numerous population of radio pulsars in globular clusters with millisecond spin periods, six slow pulsars with spin periods $0.3-4\,$s and inferred magnetic fields of $10^{11}-10^{12}\,$G have also been identified. With inferred ages of $10^8\,$yr or less, these apparently young pulsars pose a major challenge to typical neutron star formation scenarios. One possible explanation is that these objects are not young, but instead have been partially recycled \citep{VerbuntFreire2014}. In this scenario, an LMXB is interrupted by a dynamical encounter enabled by the host cluster's high central density and the mass transfer ceases before millisecond spin periods can be attained. This scenario is attractive as it naturally explains why such slow spinning pulsars are uniquely found in the densest globular clusters that have undergone or are near cluster core collapse.

However, we have demonstrated that the LMXB partial recycling scenario is not viable. In globular clusters similar to the hosts of the six slow pulsars, the typical timescale for dynamical encounters of LMXBs is roughly $10^8\,$yr or more, orders of magnitude longer than the $\lesssim 10^5\,$yr timescales required to spin up a neutron star to $1\,$s spin periods. We corroborate this conclusion using $N$-body simulations of core-collapsed clusters, finding no instances of an LMXB being dynamically interrupted before attaining a millisecond spin period.

We propose two alternative scenarios for forming slow pulsars: collapse following massive white dwarf merger \citep{Dessart2006,Schwab2021} and partial spin-up following collisions of neutron stars with luminous stars \citep{Davies1992,SigurdssonPhinney1995}. Both of these channels are expected to be most common in core-collapsed globular clusters \citep{Kremer2022,Kremer2023,Ye2024_massgap}, and thus both could naturally account for the clear preference for finding slow pulsars in the densest clusters with the highest encounter rates \citep{VerbuntFreire2014}.

Neutron stars formed via white dwarf mergers may potentially power fast radio bursts (FRBs) shortly after their formation \citep{Margalit2019}, potentially similar to the repeating FRB observed in a globular cluster in M81 \citep{Kirsten2022}. In this case, an attractive feature of the white dwarf merger scenario is it may self-consistently explain both the M81 FRB and the slow pulsars \citep{Kremer2023}. The neutron star collision scenario is unlikely to produce an FRB source because it does not provide a mechanism for sufficient magnetic field amplification. Although this connection is perhaps attractive in its simplicity, there is no reason the M81 FRB source and the slow pulsars cannot arise via separate formation channels. Future detections of additional globular cluster FRBs may further constrain this possible connection \citep{Kremer2022}.

Pulsars spun-up via accretion must occupy the space in the $P-\dot{P}$ diagram below the ``spin-up line'' indicating the shortest spin period that can be reached by accretion at the Eddington limit \citep{Pringle1972}. The fact that all six slow pulsars are found below or near the spin-up line (depending on assumptions about the uncertain accretion physics) has been touted as alternative evidence for the partial recycling scenario \citep{VerbuntFreire2014}. If slow pulsars are born via white dwarf collapse, they need not obey the spin-up line limit. Future detection of a slow pulsar lying clearly above the spin-up line would provide strong evidence that neutron star formation is indeed ongoing in globular clusters.

An additional key question concerns the incompleteness of the current sample of slow pulsars, which suffer from several selection effects. First, radio pulse beaming means that we see only roughly $1/5$ slow pulsars, compared to virtually all MSPs \citep{LyneManchester1988,Lorimer2008}. Second, long radio observations are almost always subject to red noise that causes systematically lower sensitivities to slower pulsars than expected by simply measuring the short-timescale noise levels of the data. Techniques like the Fast-Folding Algorithm \citep[e.g.,][]{Morello2020} can help mitigate this which has enabled several detections of new slow pulsars in globular clusters in the past few years \citep{Abbate2023,Zhou2024}.

Could the alternative mechanisms of Section~\ref{sec:alternatives} accommodate a true underlying population of dozens to even hundreds of slow pulsars? In the case of white dwarf mergers, the predicted population would increase by a factor of ten if neutron stars born via this channel form with weaker magnetic fields (roughly $10^{11}\,$G) allowing the pulsars to be observable for longer before evolving below the death line \citep[see Equation 4 of][]{Kremer2023}. However the magnetic field only offers so much flexibility, as field strengths $\lesssim 10^{11}\,$G would be inconsistent with the fields inferred for the current six slow pulsars \citep{Boyles2011}. Similar arguments could be made for the neutron star collision scenario, however the fine tuning argument discussed in Section~\ref{sec:NScollisions} may leave less wiggle room for this channel. The expected increase in radio pulsar detections in globular clusters over the coming years by instruments like MeerKat \citep[e.g.,][]{TRAPUM2021} and FAST \citep[e.g.,][]{FAST2021} promise to provide further insight into the full population and origin of these sources.

\acknowledgments

We thank the anonymous referee for their careful review and also thank Paulo Freire for comments on the manuscript. K.K.\ acknowledges support from the Carnegie Theoretical Astrophysics Center (CTAC), where part of this work was conducted. C.S.Y.\ acknowledges support from the Natural Sciences and Engineering Research Council of Canada (NSERC) DIS-2022-568580. F.A.R.\ acknowledges support from NSF grant AST-2108624 and NASA ATP grant 1580NSSC22K0722.

\bibliographystyle{aasjournal}
\bibliography{mybib.bib}

\end{document}